\documentstyle[12pt,epsf,epsfig]{article}
\newcommand\be{\begin{equation}}
\newcommand\ee{\end{equation}}
\newcommand\bear{\begin{eqnarray}}
\newcommand\eer{\end{eqnarray}}

\def\qsq{\vec{q}^{\,2}}
\date{ }
\begin{document}

\title{\bf $P$-Wave Polarization of the $\rho$-Meson\\
and the Dilepton Spectrum in Dense Matter}

\author{B. Friman$^a$ and H.J. Pirner$^b$\thanks{Supported in part
by the Bundesministerium f\"ur Forschung und Technologie (BMBF) under
grant no. 06 HD 742 and by GSI}\\\\
$^a$ GSI, Planckstr.~1, D-64291 Darmstadt,
Germany and\\
Institut f\"ur Kernphysik, TH Darmstadt, D-64289 Darmstadt, Germany\
\\
$^b$ Institut f\"ur Theoretische Physik der Universit\"at Heidelberg\\
Philosophenweg 19, D-69120 Heidelberg}
\vspace{1.0cm}

\setlength{\baselineskip}{24pt}

\maketitle

\begin{abstract}
We study the $p$-wave polarization operator of the $\rho$-meson due to
$\rho N$ interactions via the $N^*$ (1720) and $\Delta (1905)$
resonances and compute the corresponding production rate for
$e^+e^-$-pairs at finite temperature and baryon density. At high
baryon density we find a significant shift of the spectrum
to lower invariant masses.

\end{abstract}

\setlength{\baselineskip}{24pt}

\section{Introduction}
Heavy ion collisions at high energies have stimulated a general
interest in hadron properties at high temperature and high baryon
density.  It is expected that at sufficiently high temperature a
chiral symmetry restoring transition occurs where the constituent
quark mass becomes very
small~\cite{pis}. The temperature scale for this transition is of the
order of the Hagedorn temperature $T\simeq 160$ MeV. The equivalent
density cannot be computed reliably at present since reliable lattice
calculations of finite baryon density are not yet possible.  However,
one can perform simple estimates, e.g. by comparing the mean distance
between nucleons with their geometrical size $R_N\simeq 0.8$ fm. One
finds that the transition should take place at a baryon density 5-10
times normal nuclear matter density.

The effects of chiral symmetry restoration have been widely explored
and the spectrum of the pseudoscalar mesons has been computed within
effective chiral models
\cite{pion,kapnel,brownrhoetc,waas}. Recent
heavy-ion data from GSI are consistent, at least on a qualitative
level, with the predicted strong reduction of the K$^-$ mass at finite
baryon density \cite{senger}. For the $\pi$
meson the expected mass shift is small and does not lead to a clear
experimental signature. On the other hand, for the properties of
vector mesons in matter the situation is less clear. The theoretical
foundation for the calculations is much less firm, and different
models yield very different predictions. The additive constituent
quark model implies that lower mass constituent quarks form lower
mass bound states.  A lowering of the $\rho$ mass in matter is also
predicted by Brown and Rho using chiral symmetry and scale invariance
\cite{BR1,BR2}. On the other hand, chiral symmetry associates the
$\rho$ meson with its chiral partner the axial vector meson $a_1$,
which is higher in mass.  This may point to an increase of the
$\rho$-mass when chiral symmetry is restored and chiral partners
become degenerate~\cite{rdp}.

Neutral vector mesons mix with photons and consequently decay (with a 
small branching ratio) into lepton pairs. It is plausible that short-lived
vector mesons, like the $\rho$ meson, produced in heavy-ion collisions
decay inside the hot and dense medium. Since the lepton pairs escape
essentially without further interactions, they carry information on
the conditions in the excited hadronic matter. Recent experiments
indicate a qualitative change of the dilepton spectrum between $pA$-
and $AA$-data. Both the CERES~\cite{cer} and HELIOS-3~\cite{hel} data
show a shift of strength in the dilepton invariant mass spectrum to
lower masses, which can be interpreted in terms of a reduction of the
in-medium $\rho$ meson mass~\cite{lik,cassing1,cassing2}.

In previous calculations one considered only the self energy of a
$\rho$ meson at rest in nuclear matter. In the present paper we
discuss the possibility that the $p$-wave polarization operator
i.e. the momentum dependent self energy of the $\rho$-meson in matter,
plays an important role. For $\pi$N scattering, the (isospin
symmetric) $s$-wave interaction is known to be much less important
than that in $p$-wave states. This is partly due to chiral symmetry,
which suppresses the s-wave interaction, but also due to the existence of
the $\Delta$ resonance which is fairly close to the $\pi N$-threshold. The
$\Delta$ resonance is responsible for the strong $p$-wave polarization
of the pion in the nuclear medium. We explore the effect of the
corresponding resonances in the $\rho$N channel.

The data for $\pi N\to\rho N$ and $\gamma N\to\rho N$ 
interactions near threshold reveal two prominent resonances: the
$N^*(J=3/2^+,I=1/2)$ 1720 MeV and the $\Delta (J=5/2^+, I=3/2)$ 1905
MeV baryons. In section 2 we discuss the decay
characteristics and the resulting effective $p$-wave $\rho N$- and
$\gamma N$-couplings of these resonances. In section 3 we employ this
information to compute the properties of the $\rho$-meson in nuclear
matter. Section 4 is devoted to a calculation of the resulting
dilepton spectrum.

\section{$N^*$- and $\Delta$-resonances near the $\rho N$-
threshold}

In the review of particle properties~\cite{pdg} two
four-star baryon resonances with a large $(>50\%)$
branching ratio into the $\rho N$ channel are listed (see Table
1)\footnote{Actually D.M. Manley {\em et al.}\cite{man} need another
$3/2^+$-state at 1879 MeV to fit the $\pi N\to \pi\pi N$
data. This resonance has a large $\rho$N branching ratio 
$\Gamma_{\rho N}/\Gamma_{\rm
tot}\simeq 40\%$ in the fit of Manley {\em et al.},
but is not listed by the particle data group.}. 
\begin{table}
\begin{tabular}{lllllll}
Resonance& Mass (MeV)& $I$ & $J^P$ & $(\Gamma_{\rho N}/ \Gamma_{\rm
tot})$&$(\Gamma_{\gamma N}/\Gamma_{\rm tot})$ & $\Gamma_{\rm tot}$
(MeV)\\[0.8ex]
$N^*$ & 1720 & $1/2$ & $3/2^+$ & (70-85)\%& (0.01-0.06)\%& (100-200) \\
$\Delta$ & 1905 & $3/2$ & $5/2^+$ &  $>60\% $&(0.01-0.04)\% & (280-440) 
\end{tabular}
\caption{\em Resonance properties according to the Review of
Particle Properties.} 
\end{table}
The relative angular momentum of the $\rho$N final state is for the
$N^\star(1720)$ a pure p-wave, while for the $\Delta (1905)$ it is p-
or f-wave. For simplicity we assume that the coupling of the $\Delta
(1905)$ to the $\rho$N channel is also pure p-wave. 

\begin{figure}[hbt]
%\setlength{\unitlength}{1mm}
%\begin{picture}(150,34)
%\put(15,0){
\center{\epsfig{file=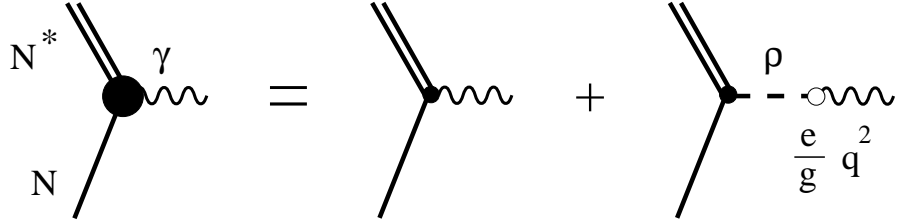,
width=30mm,angle=-90}}
%\end{picture}
\caption{The electromagnetic $N^\star N$ transition form factor in VMD.}
\label{fig:KLZ}
\end{figure}

In the calculation of the lepton-pair production rate we need the
$\gamma$N decay channel of the resonances. This is much smaller than
predicted by the Vector Meson Dominance (VMD) model of
Sakurai \cite{sakurai}. We therefore adopt the modified VMD model of
Kroll, Lee and Zumino \cite{KLZ}
\be
\label{VMD}
{\cal L}_{VMD}={e\over 2 g_{\rho\pi\pi}} F_{\mu\nu}\rho^{\mu\nu}_3,
\ee
which is explicitly gauge invariant
and allows one to fix the coupling strengths in the $\rho$N and
$\gamma$N channels independently. Here $F^{\mu\nu}$ is the 
electro-magnetic field strength tensor,
$\rho^{\mu\nu}_3 = \partial^\mu \rho^\nu_3 - \partial^\nu
\rho^\mu_3$ the tensor associated with the neutral $(i=3)$ $\rho$
meson (we retain only the abelian terms), $e$ is the electromagnetic
and $g_{\rho\pi\pi}$ the $\rho\pi\pi$ coupling
constant. For the $\gamma$N channel the
relative angular momentum is not known. We assume that the lowest
multipole dominates. The interaction
Lagrangian for the $N^*$ resonance is then of the form
\bear
\label{1} 
{\cal L}_{N^\star}&=& \frac{f_{N^\star N\rho}}{m_\rho}
\left[\Psi_{N^\star}^\dagger(\vec{S}
\times\vec{q})\cdot\vec{\rho}^{\,\,i}\tau^i \Psi_N + h.c.\right]\\ 
&+& \mu_{N^\star}
\left[\Psi_{N^\star}^\dagger(\vec S\times\vec q)\cdot\vec A\,\tau^3 \Psi_N
+ h.c. \right],
\eer 
where $\vec\rho^{\,\,i}$ is the $i$'th isospin component of the $\rho$
meson field, $\vec A$ the photon field, and $\vec S$ the
transition spin operator which connects $J=1/2$ ($N$) and $J=3/2$
($N^\star$) baryon
states \cite{sugvhip,brwei}. We note that with this Lagrangian only
transverse $\rho$ mesons can excite the $N^\star (1720)$ resonance.   
Relativistic corrections may give rise to a non-vanishing coupling for
longitudinal $\rho$ mesons. In
Fig.~\ref{fig:KLZ} the diagrams contributing to the N$^\star$N$\gamma$
form factor are shown. Note that for a real photon ($q^2 = 0$), the
$\rho$ meson contribution vanishes, due to the structure of the
$\gamma\rho$ mixing term. 
The coupling constant $f_{N^\star N\rho}$ and the transition magnetic
moment $\mu_{N^\star}$ are fixed by fitting the $\rho$N and $\gamma$N
partial widths of the $N^\star(1720)$.

The corresponding interaction Lagrangian of the $\Delta (1905)$ is more
complicated, due to the higher spin $J = 5/2$. We employ the following
form
\bear
\label{delta}
{\cal L}_\Delta&=& \frac{f_{\Delta N\rho}}{m_\rho}
\left[\Psi_{\Delta}^\dagger R_{ij} q_i \rho_j^k T^k
\Psi_N + h.c.\right]\\&+& \mu_{\Delta}
\left[\Psi_{\Delta}^\dagger R_{ij} q_i A_j\,T^3 \Psi_N
+ h.c. \right],
\eer
where $R_{ij}$ is the transition spin operator which connects $J=1/2$ and
$J=5/2$ baryon states and $T^k$ is the transition isospin operator
which connects isospin $1/2$ with isospin $3/2$ states. The transition
spin operator $R_{ij}$ is a tensor of rank 2, and a $6\times 2$ matrix
in spin space. One can construct a spin-$5/2$ object
by coupling two spin-1 objects $\epsilon^s_i$ to the nucleon
spinor $\chi^{m_s}_{1/2}$ 
\be
\label{spin-5/2}
|X^M_{ij}\rangle=\sum_{r,s,t,m_s}\,\, \Bigl({5\over 2} M \Big| 2 r
{1\over 2} m_s\Bigr)\,\, \Bigl(2 r\Big| 1 s 1 t\Bigr)
   \, \,\epsilon ^s_i \,\,\epsilon ^t_j\,\,|\chi^{m_s}_{1/2}\rangle.
\ee
The components of the unit vectors $\epsilon^s_i$ are given by
\be
\label{unit}
\epsilon^1=-{1\over 2}(1,i,0),\,\,\, \epsilon^0=(0,0,1),\,\,\, 
\epsilon^{-1}={1\over 2}(1,-i,0).
\ee
The transition spin operator is defined by
\be
\label{trans-spin}
\langle X_{ij}|N\rangle = \langle\Delta|T_{ij}
|N\rangle,
\ee
where $|\Delta\rangle$ is a six component vector
corresponding to the possible spin projections of spin-$5/2$.
As is well known for spin-$3/2$ resonances, one rarely needs to know
the explicit form of the transition spin operator. The projection operator
\be
\label{projection}
Q_{ij,kl} = \sum_{M} |X_{ij}^M\rangle\langle X_{kl}^M|
\ee
turns out to be a more useful quantity. After some tedious algebra one
finds
\bear
\label{projection2}
Q_{ij,kl} &=& {1\over 2} \left(\delta_{ik}\delta_{jl}
+\delta_{il}\delta_{jk}\right) -{1\over 5}\delta_{ij}\delta_{kl}\nonumber\\
&-&{1\over 10} \left(\delta_{ik}\sigma_j\sigma_l + \delta_{il}\sigma_j\sigma_k
+ \delta_{jk} \sigma_i\sigma_l + \delta_{jl} \sigma_i\sigma_k\right).
\eer
With the help of this operator processes involving spin-$1/2$
to spin-$5/2$ transitions are easily computed.

The partial width for the decay of the $N^\star$ resonance into the
$\rho N$ channel is given by 
\begin{eqnarray}
\Gamma_{N^\star\to N\rho}&=&\frac{2}{3}\cdot 3
\frac{f_{N^\star N\rho}^2}{m^2_\rho}\int \frac{d^4
q}{(2\pi)^4}\cdot \vec q^2\frac{m_N}{\sqrt{\vec
q^2+m^2_N}}\frac{4\pi
m_\rho\Gamma_{\rho\to\pi\pi}}{( q^2-
m^2_\rho)^2+m^2_\rho\Gamma^2_{\rho\to\pi\pi}}\nonumber\\
&&\delta(m_{N^\star}-\sqrt{\vec q^2+m_N^2}-q^0)\nonumber\\
&=&\frac{2}{\pi^2}\frac{f_{N^\star N\rho}^2}{m^2_\rho}
\int^{m_{N^\star}-m_N}_{2m_\pi} dm\frac{m_N}{m_{N^*}}\cdot m|\vec
q_1|^3\frac{m_\rho\Gamma_{\rho\to\pi\pi}}
{(m^2-m^2_\rho)^2+m^2_\rho\Gamma^2_{\rho\to\pi\pi}}
\label{resdecay}
\end{eqnarray}
where $|\vec q_1|=\left[\frac{(m^2_{N^*}-m^2_N-m^2)^2-
4m^2m^2_N}{4m^2_{N^*}} \right]^{1/2}$. Here we include the proper
phase space dependence for the $\rho$ meson decay into two pions,
\be
\label{rhopipi}
\Gamma_{\rho\to\pi\pi}(m) = \Gamma_{\rho\to\pi\pi}^{(0)}
\left(\frac{k}{k_\rho}\right)^3,
\ee
where $\Gamma_{\rho\to\pi\pi}^{(0)}= 150$ MeV,
$k = \sqrt{m^2/4 - m_\pi^2}$ and $k_R =
\sqrt{m_\rho^2/4 - m_\pi^2}$. 
By fitting the partial width
$\Gamma_{N^\star\to N\rho}\simeq 100$ MeV we find
$f_{N^\star N\rho}=7.2$, 
which is close to the coupling constant of the magnetic $\rho NN$
interaction ~\cite{bro}
\be
\label{2}
{\cal L}=i\bar
N\frac{g_{\rho\pi\pi}}{2}\frac{K^V_\rho}{2m_N}(\vec\sigma\times\vec k)
\cdot\vec\rho^{\,\, i}\tau^i N
\ee 
with $K^V_\rho=6.6\pm 0.4$, which implies $f_{N N \rho}=(m_\rho/2
m_N)(g_{\rho\pi\pi}/2)K^V_\rho = 8.1$.

For the $\gamma N$ partial width we find

\be
\label{4} 
\Gamma_{N^\star\to
N\gamma}=\frac{1}{3\pi}(\mu_{N^\star})^2
\frac{m_N}{m_{N^*}}|\vec q_2|^3
\ee 
where $|\vec
q_2|=\left(\frac{m^2_{N^*}-m^2_N}{2m_{N^*}}\right)$.
The corresponding partial widths of the $\Delta (1905)$ are given by
\be
\label{deltawidth}
\Gamma_{\Delta\to N\rho}
=\frac{1}{3\pi^2}\frac{f_{\Delta N\rho}^2}{m^2_\rho}
\int^{m_{\Delta}-m_N}_{2m_\pi} dm\frac{m_N}{m_{\Delta}}\cdot m|\vec
q_1|^3\frac{m_\rho\Gamma_{\rho\to\pi\pi}}
{(m^2-m^2_\rho)^2+m^2_\rho\Gamma^2_{\rho\to\pi\pi}}
\ee
and
\be
\label{deltagamma} 
\Gamma_{\Delta\to
N\gamma}=\frac{1}{30\pi}(\mu_{\Delta})^2
\frac{m_N}{m_{\Delta}}|\vec q_2|^3,
\ee 
where $m_{N^\star}$ should be replaced by $m_\Delta$ in $q_1$ and
$q_2$. Using $\Gamma_{\Delta\to N\rho} = 210$ MeV we find 
$f_{\Delta N\rho} = 9.0$.

The values of $\mu_{N^\star}$ and $\mu_{\Delta}$ are determined by
fitting the PDG estimates~\cite{pdg} for $\Gamma_{\gamma
N}/\Gamma_{\rm tot}$ obtained by analyzing the $\gamma N\to\pi N$
reaction.  It is convenient to introduce the ratio of the transition
magnetic moment to its value in VMD $r_{N^\star} =
\mu_{N^\star}(g_{\rho\pi\pi}/e)(m_\rho/f_{N^\star N\rho})$ and
$r_{\Delta} = \mu_{\Delta}(g_{\rho\pi\pi}/e)(m_\rho/f_{\Delta
N\rho})$. Using the maximal $\gamma N$ branching ratios we find
$r_{N^\star} \simeq 1/\sqrt{30}$ and $r_{\Delta} \simeq
1/\sqrt{5}$, respectively.  Thus the $\gamma N$ branching ratios of
the $N^\star(1720)$ and $\Delta (1905)$ resonances are suppressed by
factors of approximately 1/30 and 1/5 compared to the VMD results.

It is instructive to estimate the cross section for photoproduction of
$\rho$ mesons off protons
\bear\label{5} 
\sigma_{\gamma
N\to\rho^0 N}&\approx &\frac{\pi}{\vec k^2_{cm}}
\frac{(\Gamma_{N^\star\to\gamma N}\Gamma_{N^\star\to\rho N}) m^2_{N^\star}}{(s-
m^2_{N^\star})^2+\Gamma^2_{N^\star} m^2_{N^\star}}\nonumber\\&+& 
\frac{\pi}{\vec k^2_{cm}}\frac{3}{2}
\frac{(\Gamma_{\Delta\to\gamma N}\Gamma_{\Delta\to\rho N}) m^2_{\Delta}}{(s-
m^2_{\Delta})^2+\Gamma^2_{\Delta} m^2_{\Delta}}.
\eer 
Here $\Gamma_{N^\star}$ and $\Gamma_\Delta$ denote the full widths of
the baryon resonances.  The resulting photoproduction cross section at
low energies is smaller than the measured one~\cite{lan}.  The difference is
probably due to other partial waves, which may be described in terms
of other baryon resonances or the exchange of mesons in the
t-channel~\cite{frisoy}. Thus, the p-wave self energy of a $\rho$ meson
in nuclear matter cannot be obtained by a naive extrapolation of the
photoproduction data. Forthcoming experiments on the production of
$\rho$ and $\omega$ mesons e.g. at CEBAF will shed new light on the
mechanism for vector meson production near threshold.

\section{$\rho$N interactions at finite $\vec{q}$}

In this section we estimate the strength of the $\rho$ meson
p-wave optical potential and explore its consequences for the production of
lepton pairs in relativistic nucleus-nucleus collisions.

To lowest order in the density, the $\rho$ meson self energy in nuclear
matter is given by
\be
\label{lowdens}
\Sigma_\rho = 4 \pi f_{\rho N}\,\, \rho_B,
\ee
where $f_{\rho N}$ is the $\rho$-nucleon forward scattering amplitude
and $\rho_B$ is the baryon density. We approximate the scattering
amplitude with two resonance contributions, the $N^\star(1720)$ and
$\Delta (1905)$.  In reference \cite{HFN} a complementary approach was
chosen, where the interactions of the pions, present in a physical
$\rho$ meson, with the nuclear medium were taken into account. It was
shown that such interactions for a $\rho$ meson at rest in nuclear
matter leads to a strong increase of the width but only a small mass
shift. Similar results were obtained by Asakawa {\em et al.} \cite{AK}
and Chanfray and Schuck \cite{CS} and recently, using a different
scheme, by Klingl and Weise \cite{KW}. Here we are interested in the
momentum dependence of the self energy. A mass shift, due to other
processes like the exchange of a scalar object in the t-channel, is possible
\cite{frisoy}. 

The
$N^\star(1720)$ has a total width of 150 MeV and decays into a $\rho$ meson
and a nucleon in a relative p-wave, while
the $\Delta (1905)$ has a total width of 350 MeV and decays into the $\rho$N
channel in a p- or f-wave state \cite{pdg}. 
Since the p- and f-wave fractions of the
$\Delta (1905)\rightarrow\rho$N branching ratio are not known, we
assume, as mentioned in the section 2, that the p-wave channel
dominates and neglect the f-wave
contribution. This approximation does not significantly affect the results.
\begin{figure}[hbt]
%\setlength{\unitlength}{1mm}
%\begin{picture}(150,45)
%\put(20,0)
\center{\epsfig{file=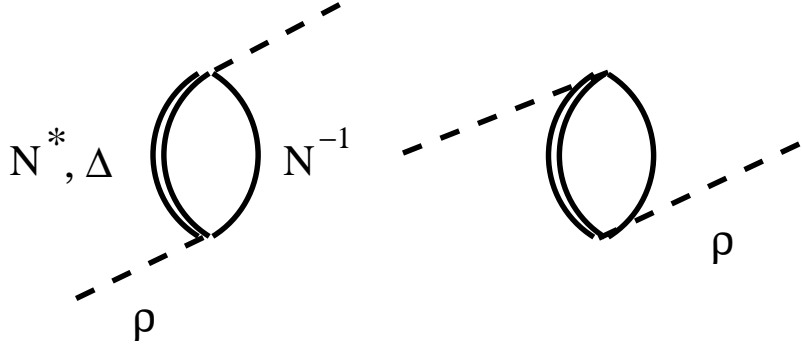,
width=45mm,angle=-90}}
%\end{picture}
\caption{The $\rho$ meson self energy diagrams in matter.}
\label{fig:p-h}
\end{figure}

The $\rho$ meson self energy in nuclear matter is now, to lowest order
in density, given by the diagrams shown if Fig.~\ref{fig:p-h}, which are easily
evaluated in the static approximation \cite{HFN}. The in-medium self
energy of transverse $\rho$ mesons due to resonance excitation is 
then given by\footnote{We consider only transverse $\rho$ mesons,
because the longitudinal ones are not strongly affected
by the p-wave self energy, since they do not couple to the $N^\star
(1720)$ in leading order.} 
\bear
\label{static}
\Sigma_\rho^{R}(\omega,\vec{q}) &=& {4\over 3}
{f_{N^\star N\rho}^2\over m_\rho^2} F(\qsq) \qsq
\rho_B{(\varepsilon_q^{N^\star} - m_N) \over \omega^2 -
(\varepsilon_q^{N^\star} - m_N)^2}\nonumber\\ &+& {2\over 5}
{f_{\Delta N\rho}^2\over m_\rho^2} F(\qsq) \qsq
\rho_B{(\varepsilon_q^{\Delta} - m_N) \over \omega^2 -
(\varepsilon_q^{\Delta} - m_N)^2},
\eer
where
\be
\label{energy}
\varepsilon_q^{N^\star} = \sqrt{\qsq + m_{N^\star}} - {i\over
2}\Gamma_{N^\star},\,\,\, \varepsilon_q^{\Delta} = 
\sqrt{\qsq + m_{\Delta}} - {i\over 2}\Gamma_{\Delta},
\ee
and $F(\qsq) = \Lambda^2/(\Lambda^2 + \qsq)$ is a form factor with 
$\Lambda = 1.5$ GeV. In eq.~(\ref{energy}) $\Gamma_{N^\star}$ and 
$\Gamma_{\Delta}$ are the
full widths, modified by
the phase space appropriate for resonances embedded in the $\rho$ meson
self energy diagrams. This point will be discussed in more detail below.
In order to be consistent with the low-density expansion, 
we employ the vacuum values for the $\rho$ meson and baryon resonance widths.
The vector-meson propagator $D_\rho(\omega,\vec{q})$ can be
represented with the help of the spectral function $A(\omega,\vec{q})$
\be
\label{prop}
D_\rho(\omega,\vec{q})=\int
d\omega^\prime\frac{A(\omega^\prime,\vec{q})}
{\omega -\omega^\prime + i \varepsilon}
\ee
\be
\label{disprel}
A(\omega,\vec{q})=
-\frac{1}{\pi}\frac{\mbox{Im}\Sigma(\omega,\vec q)}{\left[\omega^2 -\qsq
- (m_\rho)^2 - \mbox{Re}\Sigma(\omega,\vec q)\right]^2+
\left[\mbox{Im}\Sigma(\omega,\vec q)\right]^2},
\ee
where $\Sigma = - i m_\rho \Gamma_\rho + \Sigma^R_\rho$. We
retain only the imaginary part of the $\rho$ meson self energy in
vacuum. 
\begin{figure}[tb]  
%\setlength{\unitlength}{1mm}
%\begin{picture}(150,100)
%\put(20,0)
\center{\epsfig{file=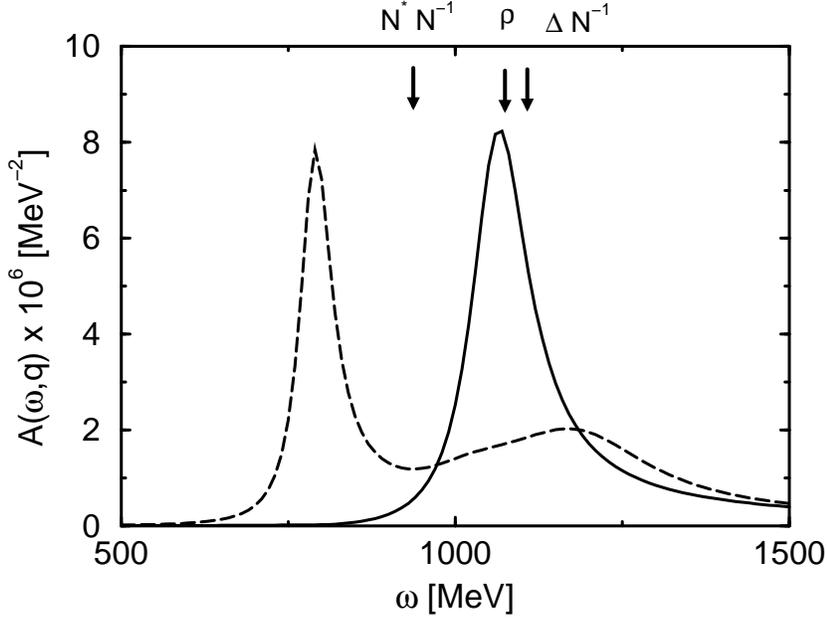,
width=100mm,angle=-90}}
%\end{picture}
\caption{The $\rho$ meson spectral function at $q = 750$ MeV in vacuum
(full line) and in nuclear matter at $\rho_B = 2 \rho_0$ (dashed
line). The arrows show the position of the unperturbed levels 
in the zero width limit.}
\label{fig:spectral}
\end{figure}
The resulting spectral function is shown in Fig.~\ref{fig:spectral} at
$q = 750$ MeV and $\rho_B = 2\rho_0$ together with the spectral
function of a $\rho$ meson in vacuum, which has an energy $\omega_q =
1.1$ GeV. Note that at finite momentum the strength in medium is moved
down to energies far below the unperturbed $N^\star N^{-1}$ and $\rho$
meson peaks. At finite
momenta one expects three states to contribute to the spectral
function, corresponding to the $\rho$ meson, and the two
resonance-hole states $N^\star N^{-1}$ and $\Delta N^{-1}$. The arrows
in Fig.~\ref{fig:spectral} indicate their unperturbed energies in the 
zero-width
limit. Due to the interactions, the levels mix, so that e.g. the
lowest level is not a pure $N^\star N^{-1}$ state, but it acquires
some $\rho$ meson strength. Furthermore, the levels are pushed apart 
(level-level repulsion), and the strength is accumulated in the
upper and lower levels, leaving very little in the middle one. This
effect, combined with the strong smearing of the spectral strength due
to the large widths of the $\rho$ meson and the baryon resonances are
responsible for the fact that there is no structure in the spectral
function at finite density corresponding to the original $\rho$ meson
peak.

We stress that it is important to take the energy dependence of the
widths properly into account. The free width of the $\rho$ meson is
completely dominated by decay into two pions. We account for the
energy dependence of this decay process by employing the standard
p-wave form (eq.~\ref{rhopipi}).
The energy dependence of the widths of the
baryon resonances, is more complicated because there are several important
decay channels. 
We assume that the phase space is
given by the decay of the resonance into the $\pi N$ channel. The
approximate treatment of the phase space is reasonable close to threshold
and is not expected to affect the results at large energies where all widths
are large anyway. Furthermore, we must express the phase space
available for the decay of a resonance embedded in a self energy
diagram in terms of the $\rho$ meson energy and momentum. This is done
in an approximate way using
\be
\label{resonance-width}
\Gamma_{N^\star}(\omega,\vec{q})= \Gamma_{N^{\star}}^{(0)}
\left(\frac{p}{p_{N^{\star}}}\right)^3, 
\ee
where
$$
p = \sqrt{\left[(s-m_N^2-m_\pi^2)^2-4 m_N^2 m_\pi^2\right]/4 s},
$$
$$
p_{N^{\star}} = \sqrt{\left[(m_{N^\star}^2-m_N^2-m_\pi^2)^2-4 m_N^2
m_\pi^2\right]/4 m_{N^\star}^2}
$$
and $\Gamma_{N^{\star}}^{(0)}=150$ MeV. 
An analogous expression is obtained for the $\Delta (1905)$. 
The invariant mass s
is computed in the static approximation for the nucleon,
$s  = (\omega+m_N)^2 - \qsq$.  
\begin{figure}[htb] 
%\setlength{\unitlength}{1mm}
%\begin{picture}(150,50)
%\put(20,8)
\center{\epsfig{file=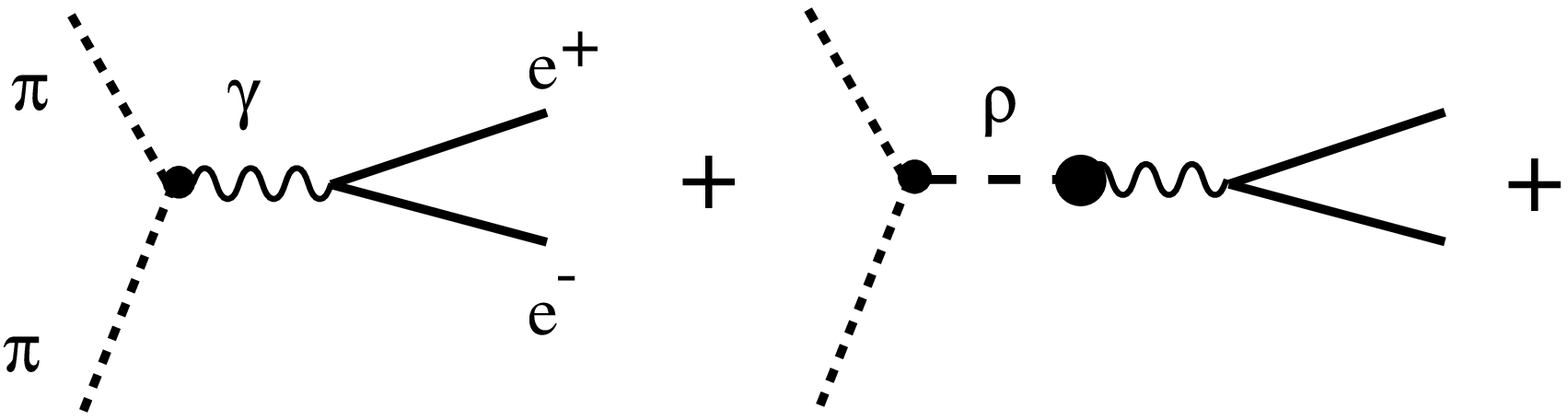,
width=35mm,angle=-90}}
%\put(40,0){\large a)}\put(100,0){\large b)}
%\end{picture}
%\begin{picture}(150,50)
%\put(10,8)
\center{\epsfig{file=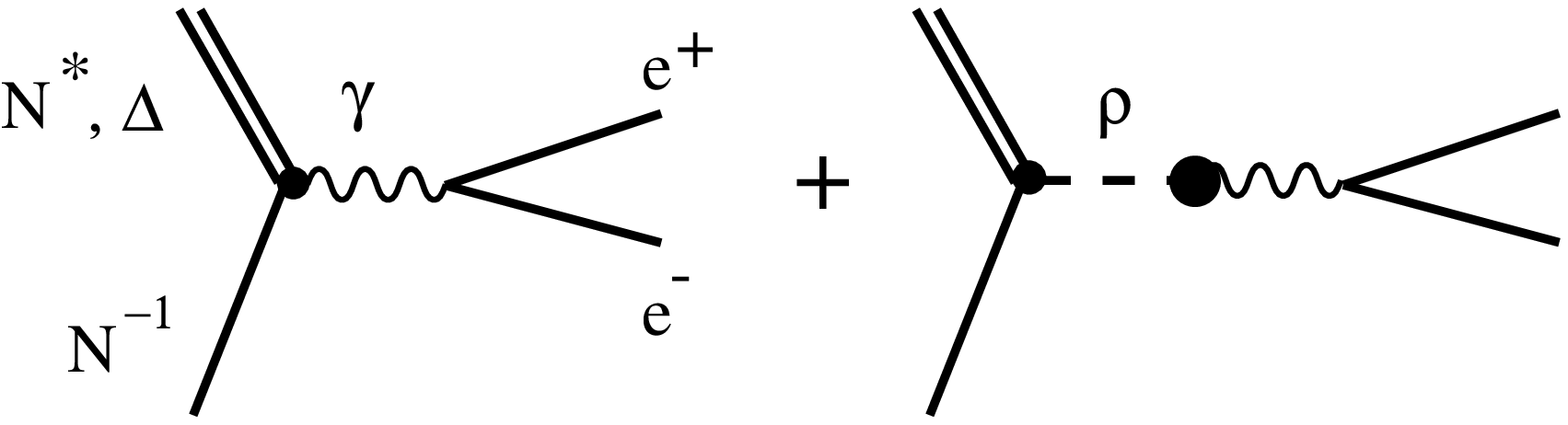,
width=35mm,angle=-90}}
%\put(40,0){\large c)}\put(100,0){\large d)}
%\end{picture}
%\begin{picture}(150,48)
%\put(40,8)
\center{\epsfig{file=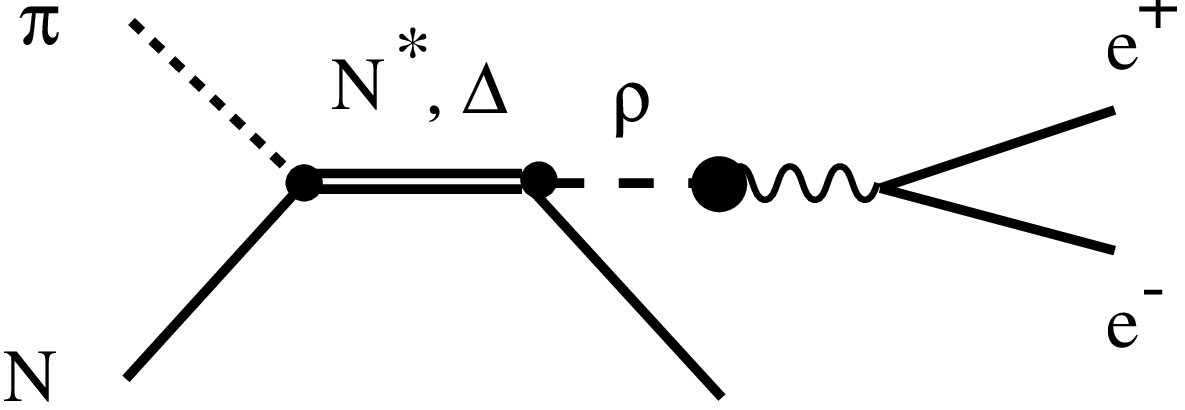,
width=30mm,angle=-90}}
%\put(70,0){\large e)}
%\end{picture}
\caption{The processes contributing to lepton pair production in the
model. The bottom diagram illustrates the scattering processes implied
by the resonance-hole contributions.}
\label{fig:dilep}
\end{figure}

The effect of the energy dependence is 
clearly seen
in the in-medium spectral function. The low lying peak is relatively
narrow because here the phase space for the resonance decays is relatively
small. On the other hand, the structures at higher energy are smeared
out by the large widths.  The low-energy peak carries about 20~\%
of the strength at $q = 750$ MeV.  As we show below, this
gives rise to a qualitative change of the lepton-pair spectrum.

The processes contributing to the $e^+e^-$ production are shown in
Fig.~\ref{fig:dilep}a-d. The corresponding rate for the production of 
$e^+e^-$ pairs per unit volume, unit time and a given invariant mass
is given by
\be
\label{rate}
\frac{d N_{e^+ e^-}}{d^3x dt dm} = \frac{8\alpha^2}{3 \pi^2 
g_{\rho\pi\pi}^2 m} 
\int_0^\infty d q \frac{q^2}{\sqrt{q^2 + m^2}} F(\sqrt{q^2 + m^2},q)
\frac{1}{\exp{\beta \sqrt{q^2+m^2}} -1},
\ee 
where $\alpha=e^2/4\pi$ and $g_{\rho\pi\pi}=6.0$ is the $\rho\pi\pi$
coupling constant. We include only the contribution
of the transverse $\rho$ mesons, which are affected by the p-wave self
energy and treat the electrons and positrons  as ultra relativistic, 
i.e., we neglect the electron mass.
The function $F(\omega,q)$ is given by
\bear
\label{rate2}
F(\omega,q)&=&
m_\rho\Gamma_\rho \left|d_\rho(\omega,q) -1\right|^2 - 
\mbox{Im}\Sigma_\rho^{N^*}(\omega,q)\left|d_\rho(\omega,q)-
r_{N^\star}\right|^2\nonumber\\
&&-\mbox{Im}\Sigma_\rho^{\Delta}(\omega,q)\left|d_\rho(\omega,q)-r_{\Delta}
\right|^2 ,
\eer
where
\be
\label{rate3}
d_\rho(\omega,q) = \frac{m^2-r_{N^\star} \Sigma_\rho^{N^*} -r_{\Delta}
\Sigma_\rho^\Delta + i m_\rho
\Gamma_\rho}{m^2-m_\rho^2-\Sigma_\rho^{N^*}-\Sigma_\rho^\Delta + i
m_\rho\Gamma_\rho}.
\ee
In the limit $r_{N^\star}=r_{\Delta}=1$ eq.~(\ref{rate2}) reduces to
the standard VMD result $F(\omega,q) = m_\rho^4 \,\,\mbox{Im}D_\rho$,
where the rate is proportional to the imaginary part
of the $\rho$ meson propagator. The realistic values of $r_{N^\star}$
and $r_{\Delta}$ are much smaller than unity (see above). 

\begin{figure}[thb] 
%\setlength{\unitlength}{1mm}
%\begin{picture}(150,25)
%\put(0,0)
\center{\epsfig{file=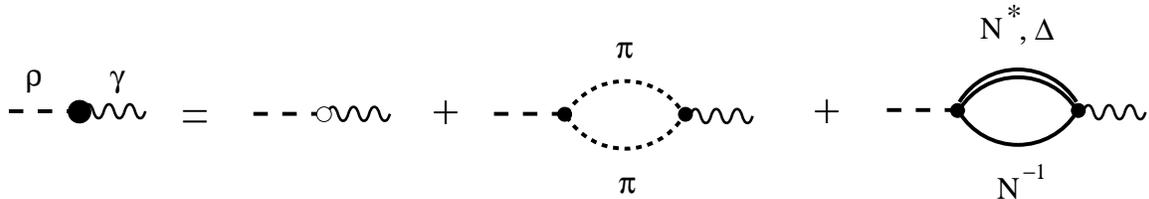,
width=25mm,angle=-90}}
%\end{picture}
\caption{Renormalization of the $\gamma\rho$ transition vertex.}
\label{fig:rhogamma}
\end{figure}
Because the resonances are unstable, the resonance-hole annihilation
diagrams {\em de facto} correspond to scattering processes, 
like the one shown in Fig.~\ref{fig:dilep}e.  Consequently, these
processes produce $e^+e^-$ pairs at all invariant masses $m> 2 m_e$.
The $\rho$ mesons in these diagrams
correspond to the full propagator, including the self energy
contributions in Fig.~\ref{fig:p-h}. For consistency the $\gamma\rho$
transition vertex is also dressed by the same diagrams (see
Fig.~\ref{fig:rhogamma}). This gives rise to the terms proportional to
$r_{N^\star}$ and $r_{\Delta}$ in the numerator of $d_\rho$. The term
proportional to $\Gamma_\rho$ is due to the two-pion loop diagram in
Fig.~\ref{fig:rhogamma}. We neglect the temperature
dependence of the self energy and the vertex renormalization. The main
effect of the finite temperature is that a fraction of
the baryons are in the form of resonances, mainly the $\Delta
(1232)$. It is difficult to correct for this, since the couplings of
these resonances to the $\rho$ meson and
higher mass baryon resonances are not known. A crude estimate
of the effect can be obtained by identifying the density in
eq.~\ref{static} with the nucleon density. This is tantamount to
setting the unknown coupling constants to zero.

\begin{figure}[htb]
%\setlength{\unitlength}{1mm}
%\begin{picture}(150,90)
%\put(13,0)
\center{\epsfig{file=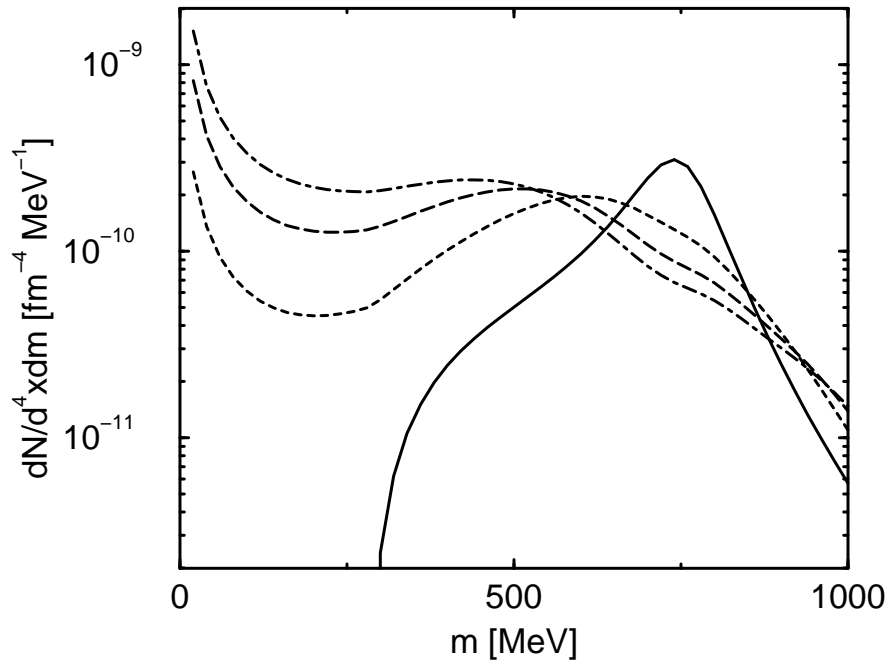,
width=100mm,angle=-90}}
%\end{picture}
\caption{The rate for lepton pair production at $T = 140$ MeV and
$\rho_B = $ 0 (full line), $\rho_0$ (dashed line), $2 \rho_0$ 
(long-dashed line) and $3 \rho_0$
(dash-dotted line).}
\label{fig:rate}
\end{figure}
In Fig.~\ref{fig:rate} the resulting production rate for $e^+e^-$
pairs is shown for various densities, at a temperature of $T = 140$
MeV. For comparison the rate due to $\pi^+\pi^-$ annihilation computed
with the $\rho$ meson propagator at $\rho_B = 0$ is also shown. The rates
include an integral over all pair momenta, but no corrections for
experimental acceptance. Clearly there is a strong enhancement of the
population of lepton pairs with invariant masses around 300-500
MeV. At finite baryon density and
finite temperature the dilepton spectrum in the mass range {\em above}
the $\rho$ resonance is modified equally due to $\pi\pi$
(figs.$4$a,b), $N^*(1720)$-hole and $\Delta$-hole (figs.$4$c,d)
annihilation. In the mass region of interest, {\em below} the $\rho$
meson peak, only the $\pi\pi$ and
$N^*(1720)$-hole self energies of the $\rho$ meson remain important.
For invariant masses below $m=400$ MeV the $N^*$-hole contribution
dominates the lepton pair spectrum. This justifies the crude
treatment of the f-wave channel in the width of the $\Delta (1905)$.

\begin{figure}[thb]
%\setlength{\unitlength}{1mm}
%\begin{picture}(150,90)
%\put(13,0)
\center{\epsfig{file=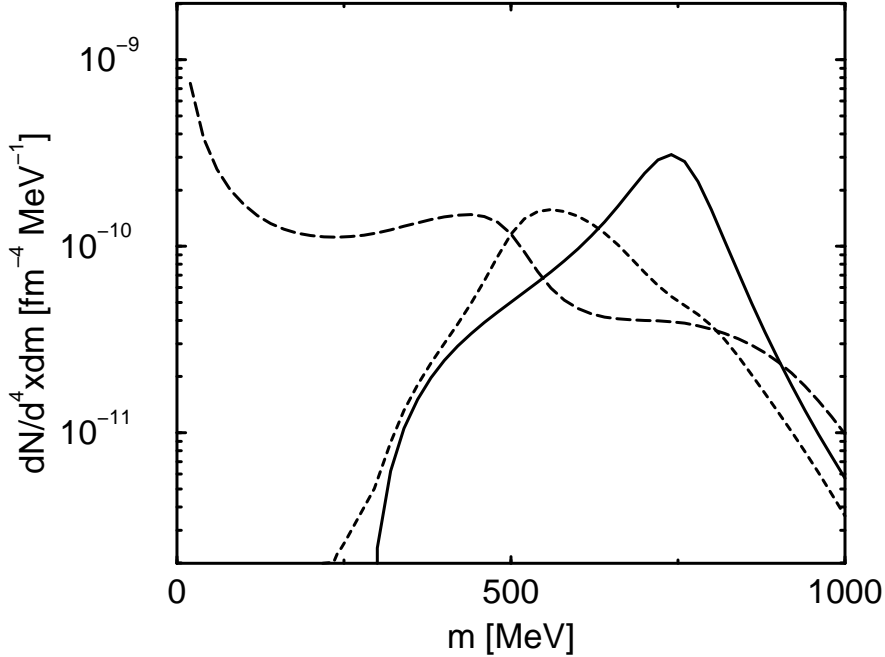,
width=100mm,angle=-90}}
%\end{picture}
\caption{The rate for lepton pair production at $T = 140$ MeV and
$\rho_B = $ 2 for momenta $q < 500$ MeV (dashed line) and $q > 500$ MeV  
(long-dashed line). For comparison the rate for $\rho_B$ = 0 is also
shown (full line).
}
\label{fig:high_low}
\end{figure}
In Fig.~\ref{fig:high_low} we show the dilepton production rate at
twice nuclear matter density and temperature $T = 140$ MeV with
cuts in the transverse momentum of the lepton pair. The $\rho
N$ interaction in the p-wave channel has a large effect on large transverse
momenta ($q > 500$ MeV), whereas the shape of the spectrum at small transverse
momenta ($q < 500$ MeV) is fairly close to that at vanishing density.
The strong momentum
dependence of the spectrum implies that a binning of the data in both
momentum and invariant mass of the lepton pairs would be a
crucial test of the p-wave $\rho$-nucleon interactions in nuclear
matter.

We have also computed the corresponding photon production rate, which
is due only to the resonance-hole annihilation processes. The so
obtained photon production rate is
comparable to the rates of the dominating processes
($\pi\pi\rightarrow\rho\gamma$ and $\pi\rho\rightarrow \pi\gamma$) in
the calculation of Kapusta {\em et al.}~\cite{KLS} at photon energies
between 0.5 and 1 GeV and far below those rates at higher
energies. Consequently, we do not expect the p-wave mechanism to be in
conflict with the photon spectra in heavy-ion collisions.

\section{Discussion and conclusions}

We have computed the low mass enhancement of the lepton-pair
production rate due to $\rho N$ interactions, where the nucleon is
excited into the $N^\star (1720)$ and $\Delta (1905)$ resonances. 
The $N^*(1720)$ resonance is very
clearly seen in $\pi N\rightarrow \pi\pi N $ reactions with a resonant
$\rho(\pi\pi)$ final states.  Furthermore, the mass of the $N^\star
(1720)$ is close to the sum of the $\rho$ meson and nucleon masses.
Therefore, it seems natural that an
important contribution to the $\rho$ meson self energy in baryonic
matter is due to this resonance. 

We note that the calculation presented here does not address
the problem of low mass dileptons at finite temperature and vanishing 
baryon density. In this case the interactions with the surrounding mesons
\cite {haglin} are important.  The time scale set by the dilepton
invariant mass $(\tau \sim m^{-1})$ allows the quark antiquark pair to
probe long distance interactions. Thus, low-mass lepton pairs explore
non-perturbative gluon configurations at high
temperatures. Because of the high density of hadrons in
heavy-ion collisions at ultra-relativistic energies the studies within
hadronic models should 
definitely be supplemented by further analysis in terms of quark and
gluon degrees of freedom. Dynamical properties of constituent
quarks at finite temperature may be reflected in the lepton pair
spectrum \cite{PW}.  

The properties of $\rho$ mesons in matter can also be studied with
QCD sum rules \cite{HL}. Recently, this approach has been extended to
$\rho$ mesons at finite momentum \cite{FL}. Preliminary results
indicate that the p-wave transverse $\rho$-polarization operator is
attractive as in our calculation, and furthermore that the
longitudinal self energy is weakly repulsive.

The experimentally observed low mass enhancement
cannot be reproduced by transport models without modifications of the
$\rho$ meson properties at high temperatures and densities. 
In this paper we consider the production rate for lepton pairs in a
high temperature high density system in thermodynamic equilibrium. 
For a direct comparison
with experiment a realistic description of the collision dynamics, as
given e.g. by a transport simulation of nucleus-nucleus collisions, is
needed. In recent work, Rapp, Chanfray and Wambach \cite{CRW}
have included the p-wave polarization of the $\rho$ meson discussed
here in their dynamical model. They find quantitative agreement with
the CERES data. A substantial part of the low mass enhancement is in
their model due to the p-wave mechanism. 

The CERES and HELIOS-3 data can also be interpreted in terms of a
model with reduced vector meson masses in matter \cite
{lik,cassing1,cassing2,SB}.  There is an important difference between
such a model with an s-wave self energy and our model with a p-wave
$\rho$ meson self energy.  The differential cross section for lepton
pairs at large transverse momenta is strongly enhanced by the p-wave
self energy, while the cross section at small transverse momenta is
almost unchanged.  This is a distinctive feature which allows one to
experimentally discriminate between the two models. In reality
probably both medium effects will be present, but the transverse
momentum dependence can be used to pin down the relative importance of
the s- and p-wave contributions to the $\rho$ meson self energy in
nuclear matter.

\begin{figure}[thb]
%\setlength{\unitlength}{1mm}
%\begin{picture}(150,90)
%\put(13,0)
\center{\epsfig{file=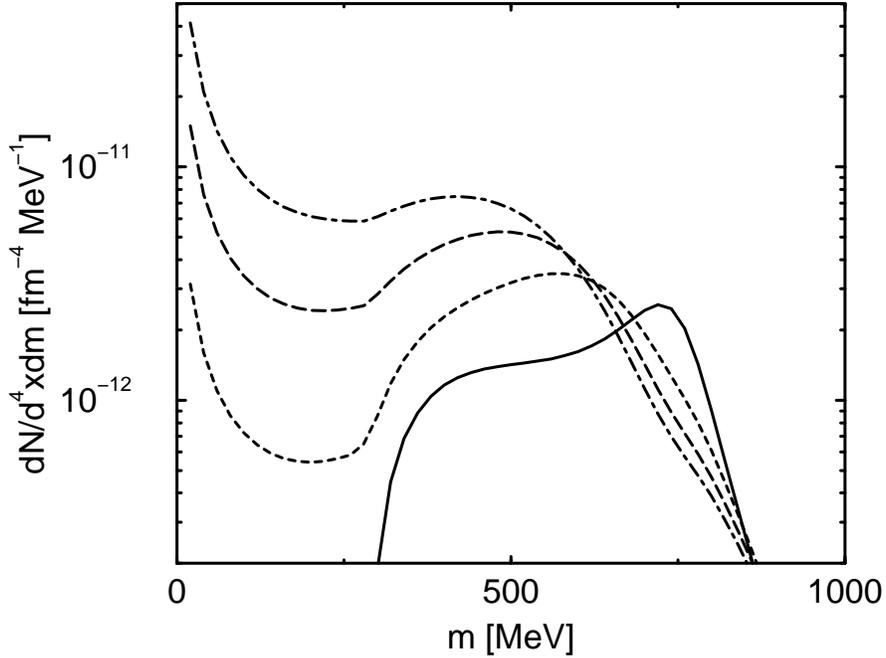,
width=100mm,angle=-90}}
%\end{picture}
\caption{Same as Fig.~6 but for $T = 80$ MeV.
}
\label{fig:low_T}
\end{figure}
We note that the relative enhancement of the low mass lepton pair
spectrum due to the p-wave mechanism is stronger at lower
temperatures. In Fig.~\ref{fig:low_T} we show the production rate at
$T=80$ MeV, which together with a density in the range 2-3 $\rho_0$
represents the conditions probed in heavy-ion collisions at GSI
energies (1-2 GeV/A). Because the temperatures are moderate, these
experiments explore the properties of matter where the baryon
density plays a key role. For such a system, the approximation where
we neglect the temperature dependence of the polarization operator is
a reasonable first approximation. Thus, the future dilepton
experiment at GSI, HADES, will be sensitive not only to shifts of the $\rho$
meson mass, but also to the $\rho$ meson p-wave self energy in dense
matter.  Forthcoming experimental efforts both at GSI and CERN will
provide more accurate data, which may pin down the mechanism for
medium modifications of vector mesons in matter. Complementary
information on the properties of vector mesons in cold matter can be
obtained by studying $\rho$ meson production off nuclei with
photon and hadron beams.

\section*{Acknowledgment}
We are grateful to W. Cassing, A. Drees, V. Koch and K. Redlich for
valuable discussions.


\begin{thebibliography}{WWW}
\bibitem{pis} R.D.~Pisarski, Phys. Lett. {110B} (1982) 155.
\bibitem{pion} H.~Meyers-Ortmanns, H.J.~Pirner, B.-J.~Schaefer,
Phys. Lett. B311 (1993) 213;
H. Meyer-Ortmanns, B.-J. Schaefer, Phys.Rev.D53 (1996) 6586.
\bibitem{kapnel} D.B.~Kaplan and A.E.~Nelson, Phys. Lett. B175 (1986) 57.
\bibitem{brownrhoetc} G.E.~Brown, K.~Kubodera, M.~Rho and V.~Thorsson,
Phys. Lett. B291 (1992) 355; G.E.~Brown, C.H.~Lee, M.~Rho and
V.~Thorsson, Nucl. Phys. A567 (1994) 937.
\bibitem{waas} T.~Waas, N.~Kaiser and W.~Weise, Phys. Lett. B365
(1996) 12; Phys. Lett. B379 (1996) 34.
\bibitem{senger} P.~Senger, in Proc. Strangeness '96, Budapest, May,
1996 (to be published in Heavy Ion Physics); Proc. Meson '96, Cracow,
May 1996 (to be published in Acta Phys. Pol.).
\bibitem{BR1} G.E.~Brown and M.~Rho, Phys. Rev Lett. {66} (1991)
2720.
\bibitem{BR2} G.E.~Brown and M.~Rho, Phys. Reports {269} (1996)
333.
\bibitem{rdp} R.D.~Pisarski, Phys. Rev. {D52} (1995) 3773-3776;
Nucl. Phys. {A590} (1995) 553c;
Proc. Int. Workshop on Nuclear and Particle Physics: {Chiral
dynamics in hadrons and nuclei}, February 6-11, 1995, Seoul, Korea.
\bibitem{cer} CERES, G.~Agakichiev et al.,  Phys. Rev. Lett.
{75} (1995) 1272; Th. Ullrich, Proc. of Quark Matter '96, Heidelberg,
1996, to appear in Nucl. Phys. A.
in these proceedings.
\bibitem{hel} HELIOS-3, M.~Masera {\em et al.}, Nucl. Phys. {A590}
(1992) 93c.
\bibitem{lik} G.Q. Li, C.M. Ko, and G.E. Brown, Phys. Rev. Lett.
{75} (1995) 4007 and Nulc. Phys. {A606} (1996) 568.
\bibitem{cassing1} W. Cassing, W. Ehehalt and C.M. Ko, Phys. Lett.
{B363} (1995) 35
\bibitem{cassing2} W. Cassing, W. Ehehalt and I. Kralik, Phys. Lett.
{B377} (1996) 5.
\bibitem{pdg} L.~Montanet {\em et al.},
Review of particle properties, Particle Data Group,
Phys. Rev. {D50} (1994) 1173-1823.
\bibitem{man} D.M.~Manley and E.M.~Saleski, Phys. Rev. {\bf D45}
(1992) 4002.
\bibitem{sakurai} J.J.~Sakurai, Currents and mesons (Univ. of Chicago
Press, Chicago, 1969); Ann. Phys. 11 (1960) 1. 
\bibitem{KLZ} N.M.~Kroll, T.D.~Lee and B.~Zumino, Phys. Rev. 157
(1967) 1376.
\bibitem{sugvhip} H.~Sugawara and F.~von Hippel, Phys. Rev. 172 (1968) 1764.
\bibitem{brwei} G.E.~Brown and W.~Weise, Phys. Reports C22 (1975) 279.
\bibitem{bro} G.E.~Brown, Nucl. Phys. {\bf A446} (1985)  10c
\bibitem{lan} Hanbook of Physics, Landolt-B\"ornstein, Photoproduction
of elementary particles, Vol. 8, p. 322.
\bibitem{frisoy} B.~Friman and M.~Soyeur, Nucl. Phys. {A600} (1996) 477.
\bibitem{HFN} M.~Herrmann, B.~Friman, and W.~N\"orenberg, Nucl. Phys.
{A560} (1993) 411.
\bibitem{AK} M.~Asakawa {\em et al.}, Phys. Rev. C46 (1992) R1159.
\bibitem{CS} G.~Chanfray and P.~Schuck, Nucl. Phys. A545 (1992) 271c.
\bibitem{KW} F.~Klingl and W.~Weise, Nucl. Phys. A606 (1996) 329
\bibitem{KLS} J.~Kapusta, P.~Lichard and D.~Seibert, Phys. Rev. D 44
(1991) 2774; Phys. Rev. D 47 (1993) 4171 
\bibitem{haglin} K.~Haglin, Phys. Rev. {C53} (1996) 2606.
\bibitem{PW} H.J.~Pirner and M.~Wachs, submitted to Nucl. Phys. A.
\bibitem{HL} T.~Hatsuda and S.-H.~Lee, Phys. Rev. {C46} (1992) 34.
\bibitem{FL} B.~Friman and S.H.~Lee, in preparation.
\bibitem{CRW} G.~Chanfray, R.~Rapp and J.~Wambach, to be published.
\bibitem{SB} H.J.~Schulze, D.~Blaschke, Phys. Lett. B386 (1996) 429.
\end{thebibliography}
\end{document}